\documentclass[aps,prl,superscriptaddress,showpacs,twocolumn]{revtex4}
\usepackage{graphicx}
\usepackage{epsfig,bbm}
\newcommand{\be}{\begin{equation}}
\newcommand{\ee}{  \end{equation}}
\newcommand{\ba}{\begin{eqnarray}}
\newcommand{\ea}{  \end{eqnarray}}

\begin{document}
\title{Chaotic Scattering in the Regime of Weakly Overlapping Resonances}
\author{B.~Dietz}
\affiliation{Institut f\"{u}r Kernphysik, Technische
Universit\"{a}t Darmstadt, D-64289 Darmstadt, Germany}
\author{T.~Friedrich}
\affiliation{Institut f\"{u}r Kernphysik, Technische
Universit\"{a}t Darmstadt, D-64289 Darmstadt, Germany}
\author{H.~L.~Harney}
\affiliation{Max-Planck-Institut f\"{u}r Kernphysik, D-69029
Heidelberg, Germany}
\author{M.~Miski-Oglu}
\affiliation{Institut f\"{u}r Kernphysik, Technische
Universit\"{a}t Darmstadt, D-64289 Darmstadt, Germany}
\author{A.~Richter}
\affiliation{Institut f\"{u}r Kernphysik, Technische
Universit\"{a}t Darmstadt, D-64289 Darmstadt, Germany}
\author{F.~Sch\"afer}
\affiliation{Institut f\"{u}r Kernphysik, Technische
Universit\"{a}t Darmstadt, D-64289 Darmstadt, Germany}
\author{H.~A.~Weidenm\"{u}ller}
\affiliation{Max-Planck-Institut f\"{u}r Kernphysik, D-69029
Heidelberg, Germany}
\date{\today}

\begin{abstract}
We measure the transmission and reflection amplitudes of
microwaves in a resonator coupled to two antennas at room
temperature in the regime of weakly overlapping resonances and in
a frequency range of $3$ to $16$ GHz. Below $10.1$ GHz the
resonator simulates a chaotic quantum system. The distribution of
the elements of the scattering matrix $S$ is not Gaussian. The
Fourier coefficients of $S$ are used for a best fit of the
autocorrelation function of $S$ to a theoretical expression based
on random--matrix theory. We find very good agreement below but
not above $10.1$ GHz.
\end{abstract}

\maketitle Chaotic quantum scattering occurs when Schr{\"o}dinger
waves are scattered by a system with chaotic classical dynamics.
For time--reversal invariant chaotic systems, the spectral
fluctuations of the eigenvalues coincide~\cite{Boh84} with the
predictions of the Gaussian Orthogonal Ensemble (GOE) of real and
symmetric random matrices. The eigenvalues manifest themselves as
resonances with average spacing $D$ and average width $\Gamma$.
The theory of chaotic scattering has been largely developed in the
framework of nuclear reaction theory~\cite{Bro81}. Predictions of
the theory have been thoroughly tested both in the regime of
isolated resonances ($\Gamma \ll D$)~\cite{Lyn68} and in the
Ericson regime ($\Gamma \gg D$)~\cite{Eri60}, especially in the
context of nuclear physics~\cite{Eri63} but also in several other
areas of physics~\cite{Blu88}. In contradistinction, we are not
aware of any thorough investigation of chaotic scattering in the
regime of weakly overlapping resonances that would comprise all complex 
reflection and transmission elements of the scattering matrix.
In this Letter, we present data in that regime and compare
these with theoretical predictions.

\begin{figure}[!hbt]
\epsfig{figure=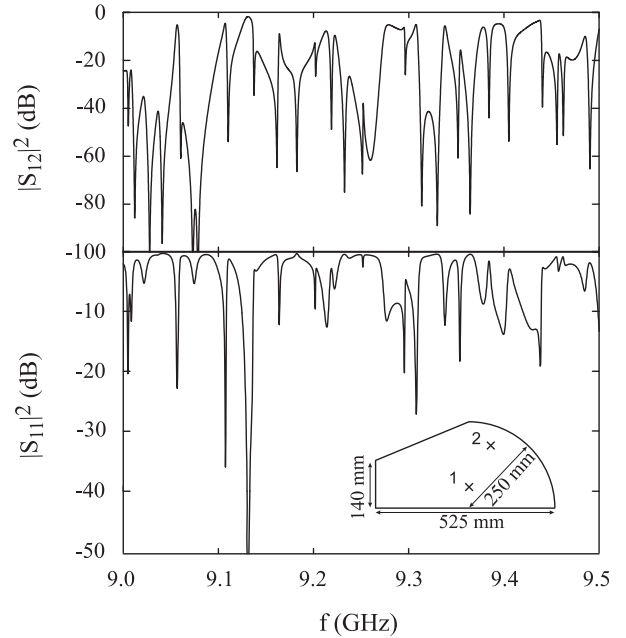,angle=0,width=8cm} \caption{Absolute
squares of the scattering matrix elements $S_{a b}$ for signal
transmission from antenna 2~to~1 (upper panel) and reflection at
antenna 1 (lower panel) between 9.0 and 9.5~GHz. On the
logarithmic decibel scale $-x$ dB means an attenuation of the
microwave power by the factor $10^{x/10}$. The resonances overlap
and create a pattern of fluctuations. Insert: The shape of the
two--dimensional microwave resonator used in the experiment. The
points 1 and 2 indicate the positions of the antennas.}
\label{fig1}
\end{figure}

{\it Experiment}. We use a microwave cavity made of Copper coupled
to two antennas and measure the response to an external field as a
function of radiofrequency $f$. The microwave cavity has the shape
of a tilted stadium billiard~\cite{Pri94}, see the insert of
Fig.~\ref{fig1}. The dynamics of the classical stadium billiard is
chaotic. The tilted shape was used in order to avoid
bouncing--ball orbits between parallel walls. The height of the
cavity is 14.6~mm. For frequencies $f \leq f_{\rm max} =
10.1$~GHz, only a single vertical mode in the microwave cavity is
excited. In that regime, the cavity simulates a two--dimensional
chaotic quantum system and is a microwave billiard~\cite{Sto90}.
The experiment is performed at room temperature, with Ohmic losses
at the walls of the cavity.
A vector network analyzer couples microwave power in and out of
the resonator via either one or both antennas and yields the
complex elements $S_{ab}(f)$ of the symmetric scattering matrix,
where $a, b = 1, 2$.  The range 3~GHz $\,\le f\le\,$16~GHz was
covered in steps of $\Delta = $~250~kHz in reflection measurements
(yielding $S_{11}(f)$ and $S_{22}(f)$) and of $\Delta=$~100~kHz in
transmission measurements (yielding $S_{12}(f)$). Fig.~\ref{fig1}
gives examples of the measured transmission and reflection
intensities.

Fig.~\ref{fig2} shows histograms of the distribution of
$S$--matrix elements in two frequency intervals. The distribution
of ${\rm Re}\{S_{1 1}\}$ is strongly peaked near $1$, especially
for the lower interval, and obviously not Gaussian. The
distributions of ${\rm Im}\{S_{1 1}\}$ and of ${\rm Re}\{S_{1
2}\}$ deviate from Gaussians (solid lines). The distributions of
the phases (rightmost panels) are peaked.

We use the data to construct the $S$--matrix autocorrelation
functions $C_{ab}(\epsilon) = \overline{S_{ab}(f)S^\ast_{ab}(f +
\epsilon)} - | \ \overline{S_{ab}(f)} \ |^2$ for $a,b=1,2$. The
bar denotes an average over a frequency window. Three examples for
$C_{ab}(\epsilon)$ are displayed as points in the two upper panels
of Fig.~\ref{fig3} (data taken below $f_{\rm max} = 10.1$~GHz),
and in the insert of Fig.~\ref{fig4} (data from above $f_{\rm
max}$ where the cavity does not simulate a two--dimensional
microwave billiard). The values of the scattering matrix $S_{a
b}(f)$ are seen to be correlated, with a correlation width $\Gamma
\approx$ several MHz. With $S^{\rm fl}_{a b} = S_{a b} -
\overline{S_{a b}}$ we have also determined the ``elastic
enhancement factor'' $W = \bigg( \overline{|S^{\rm fl}_{1 1}|^2} \
\overline{|S^{\rm fl}_{2 2}|^2} \bigg)^{1/2} / \ \overline{|S^{\rm
fl}_{1 2}|^2}$ as a function of $f$, both from the autocorrelation
functions and from the widths of the distributions of the
imaginary parts of the scattering matrix (Fig.~\ref{fig2}). Both
results agree very well and yield a smooth decrease of $W$ with
$f$ from $W \approx 3.5\pm 0.7$ for $4 \leq f \leq 5$ GHz to $W \approx
2.0\pm 0.7$ for $9 \leq f \leq 10$ GHz. The computation of the
enhancement factors based on a theoretical expression for the S-matrix
autocorrelation function introduced below yields the values W=2.8 and
W=2.2, respectively. Moreover, we have converted the scattering
functions $S_{a b}(f)$ (measured at $M$ equidistant frequencies with
step width $\Delta$) into complex Fourier coefficients $\tilde{S}_{a
b}(t)$ with $t \geq 0$. Instead of the Fourier index $k$ we use the
discrete time interval $t = k / (M \Delta)$ elapsed after excitation
of the resonator. The Fourier coefficient $\tilde{S}_{a b}(0)$ is
proportional to $\overline{S_{a b}(f)}$. We find that
$\tilde{S}_{1 2}(0) \approx 0$. Any two complex Fourier coefficients $\tilde{S}_{a b}(t)$ of
$S_{a b}(f)$ are uncorrelated random variables~\cite{Eri65}. For $t >
0$, the coefficients $\tilde{S}_{a b}(t)$ have an approximately
Gaussian distribution about their ($t$--dependent) mean
value~\cite{footnote}.  This result is unexpected and was neither predicted theoretically
nor found experimentally before. 

\begin{widetext}\onecolumngrid
\begin{figure}[b!]
\epsfig{figure=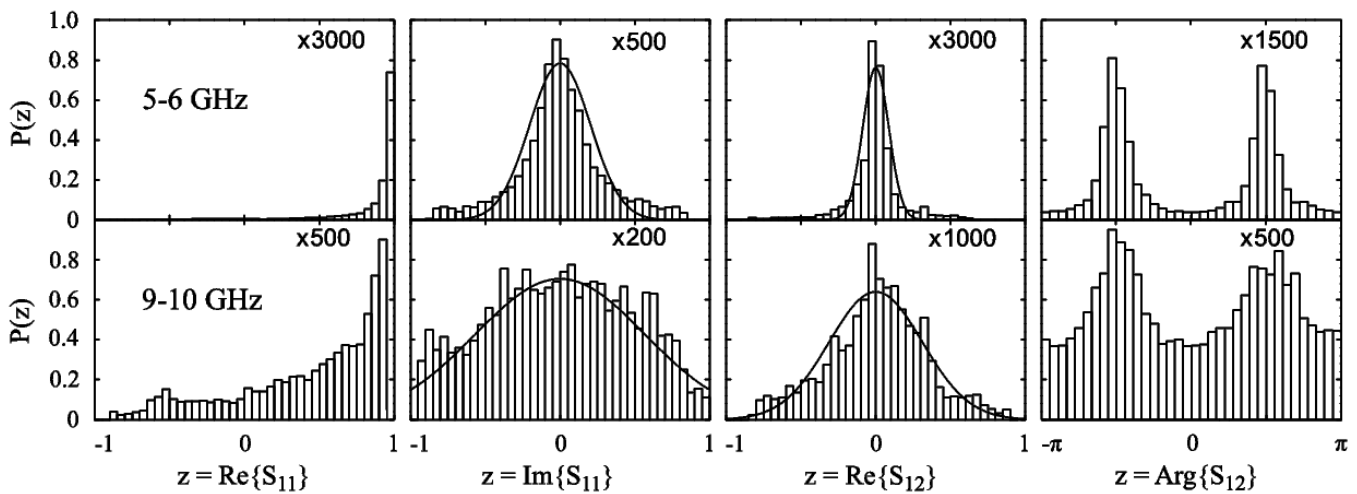,angle=0,width=\linewidth} \caption{From
left to right: Histograms for the scaled distributions of the real
and imaginary parts of the reflection amplitude $S_{1 1}$ and the
real part and the phase of the transmission amplitude $S_{1 2}$,
for the two frequency intervals $5$--$6$ GHz (upper panels) and
$9$--$10$ GHz (lower panels). The scaling factors are given in
each panel. The solid lines are best fits to Gaussian
distributions.} \label{fig2}
\end{figure}
\end{widetext}

The Fourier transform $\tilde{C}_{a b}(t)$ of $C_{a b}(\epsilon)$
has Fourier coefficients $x_t = |\tilde{S}_{a b}(t)|^2$. In the
lower panels of Fig.~\ref{fig3} (in Fig.~\ref{fig4}) we show data
for $\log_{10} \tilde{C}_{a b}(t)$ versus $t$ for two values of
$\{a, b\}$ (for $\{a, b\} = \{1, 2\}$, respectively). The cutoff
at $t = 800$ ns in both figures is due to noise. The $\tilde{S}_{a
b}(t)$ being nearly Gaussian, the distribution $P(y_t)$ of $y_t =
\ln x_t$ is expected to have approximately the form
\begin{equation}
P(y_t) = \exp{(y_t - \eta_t - e^{y_t - \eta_t})} \label{vertlog}
\end{equation}
where $\eta_t=\ln{\bar{x}_t}$ is given by the expectation value of
$x_t$. The maximum of $P(y_t)$ is at $y_t = \eta_t$, and $P(y_t)$
has a strong skewness due to the exponential within the argument
of the exponential, in agreement with the experimental data.

\begin{figure}[!hbt]
\epsfig{figure=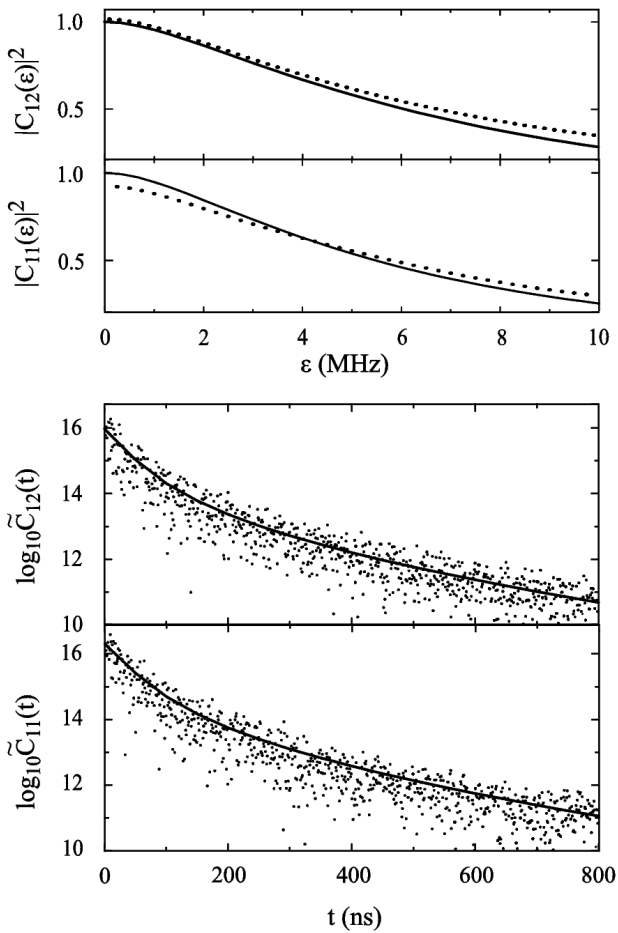,angle=0,width=8.3cm} \caption{Upper
panels: Comparison of the autocorrelation function $C_{a
b}(\epsilon)$ constructed from the data (points) and the fit using
Eq.~(\ref{vwz_formel}) (full line), both normalized by the value
of $C(0)$ as given by Eq.~(\ref{vwz_formel}). Lower panels:
Fourier coefficients $\tilde{C}_{a b}(t)$ of the autocorrelation
functions (points) and the Fourier transform of the fit of $C_{a
b}(\epsilon)$ as given by Eq.~(\ref{vwz_formel}) to the data (full
line). The elements $S_{1 2}(f)$ and $S_{1 1}(f)$ were taken from
the frequency window 9--10~GHz.} \label{fig3}
\end{figure}
\begin{figure}[!hbt]
\epsfig{figure=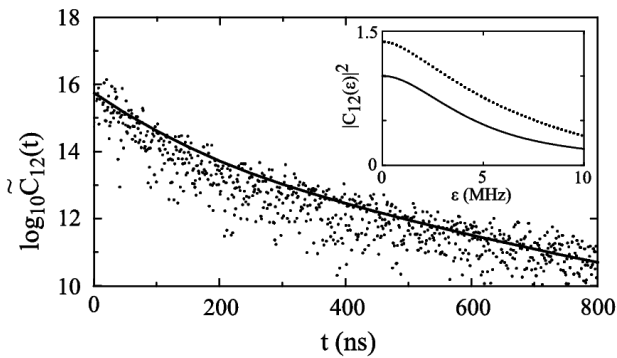,angle=0,width=8.3cm} \caption{Same as
Fig.~\ref{fig3} but for the scattering function $S_{1 2}(f)$ taken
in the frequency window 12--13~GHz.} \label{fig4}
\end{figure}

{\it Theory}. In the regime of weakly overlapping resonances, the
only theory available is due to Verbaarschot, Weidenm{\"u}ller,
and Zirnbauer~\cite{Ver85} (in the sequel: VWZ). These authors
model the scattering matrix $S$ of a time--reversal invariant
system in terms of a GOE Hamiltonian matrix of dimension $N$. In
the absence of ``direct reactions'' (i.e., for $\overline{S_{1
2}(E)} = 0$), the relevant parameters of the theory are the
``transmission coefficients'' $T_c = 1-| \ \overline{S_{cc}(f)} \
|^2$ which measure the unitarity deficit of the average
$S$--matrix. Given the $T_c$, the theory uses the limit $N \to
\infty$ to predict for all values of $\Gamma / D$ the $S$--matrix
autocorrelation function
\begin{eqnarray}
\label{vwz_formel} &&C_{ab}(\epsilon) = \frac{1}{8}
\int\limits_0^\infty d\lambda_1 d\lambda_2 \int\limits_0^1
d\lambda\, \mu(\lambda, \lambda_1, \lambda_2)
J_{ab}(\lambda, \lambda_1, \lambda_2) \nonumber \\
&& \times \exp(-i \pi \epsilon (\lambda_1 + \lambda_2 +
2\lambda)/D)
\nonumber \\
&& \times \prod\limits_c \frac{(1-T_c\lambda)}{((1+T_c\lambda_1)
(1+T_c\lambda_2))^{1/2}}
\end{eqnarray}
in terms of the ratio $\epsilon / D$. To simulate Ohmic absorption
by the walls of the cavity, we introduce additional fictitious
channels~\cite{Sch03} and associated transmission coefficients
$T_c$ with $c = 3, 4, \ldots$. These are defined below. The
product over channels $c$ extends over both, the antenna channels
and the ficticious channels. The function $J_{ab}(\lambda,
\lambda_1, \lambda_2)$ depends on the $\lambda$s and on the
transmission coefficients $T_a$, $T_b$ for the open channels. Both
the integration measure $\mu(\lambda, \lambda_1, \lambda_2)$ and
$J_{a b}$ are given explicitly in Ref.~\cite{Ver85}. The
correlation width $\Gamma$ is actually determined by
Eq.~(\ref{vwz_formel}) but approximately given by the ``Weisskopf
estimate'' $\Gamma \approx [ D / (2 \pi) ] \sum_c T_c$.
Equation~(\ref{vwz_formel}) comprises what is known theoretically
in the regime of weakly overlapping resonances. Higher moments of
$S$ are not known, not to speak of the complete distribution of
$S$--matrix elements.

Much more is known both for $\Gamma \gg D$ and for $\Gamma \ll D$.
In the Ericson regime, the distribution of $S$--matrix elements is
Gaussian; the correlation function $C_{a b}(\epsilon)$ has
Lorentzian shape, with $\Gamma$ given by the Weisskopf
estimate~\cite{Aga75}; the Fourier transform $\tilde{C}_{a b}(t)$
of $C_{a b}(\epsilon)$ (which describes the decay in time of the
modes in the cavity) is exponential in time; for $T_c \approx 1$
(strong absorption) the distribution of the phases of the $S_{a
b}$ is constant. For $\Gamma \ll D$, on the other hand, the
distribution is far from Gaussian. (Consider, f.i., the
single--channel case. The unitarity condition $|S(f)| = 1$
confines $S(f)$ to the unit circle. The phase of $S(f)$ increases
by $2 \pi$ over the width of every resonance and is nearly
stationary in between resonances.) The regime $\Gamma \approx D$
interpolates between these two extremes and we expect a
non--Gaussian distribution of $S(f)$. The results in
Fig.~\ref{fig2} give experimental information on that distribution
and confirm our expectation. With decreasing $f$, the
distributions deviate ever more strongly from Gaussians. As for
$\tilde{C}_{a b}(t)$, Eq.~(\ref{vwz_formel}) predicts a
power--like decay in time, in striking contrast to the exponential
decay valid for $\Gamma \gg D$. That prediction has been discussed
and used in Refs.~\cite{vwz_theo} and experimentally tested with
microwave resonators in Refs.~\cite{Sch03,vwz_exp}. However, these
papers did not apply any statistical tests based upon a
goodness-of-fit (GOF) as done below.

{\it Analysis}. We model Ohmic absorption by a large number of
absorptive channels with very small transmission coefficient
each~\cite{Sch03}. The product in Eq.~(\ref{vwz_formel}) over
absorptive channels is then replaced by an exponential function of
the sum $\tau_{\mathrm{abs}}$ of the transmission coefficients of
these channels, and $C_{a b}(\epsilon)$ depends on $T_1, T_2,
\tau_{\mathrm{abs}}$ and $D$. The Fourier transform was fitted to
the $x_t$--data shown in the lower parts of Figs.~\ref{fig3}
and~\ref{fig4}. We used $T_a = 1 - | \ \overline{S_{a a}} \ |^2$
for $a = 1,2$ and calculated the mean level spacing $D$ from the
Weyl formula~\cite{Bal76}. This left $\tau_{\mathrm{abs}}$ as the
only free parameter. In order to allow for secular variations of
$\tau_{\mathrm{abs}}$, the data taken between 3 and 16~GHz were
analyzed in 1~GHz intervals with the help of a maximum likelihood
fit. We find that the sum $T_1 + T_2 + \tau_{\mathrm{abs}}$
increases from 0.11 in the interval 3--4~GHz to 1.15 in the
interval 9--10~GHz. The resulting increase of
$\tau_{\mathrm{abs}}$ is consistent with conductance properties of
Copper. Using the Weisskopf estimate we find that $\Gamma/D$
increases from 0.02 to 0.2 over the same range.  This shows that
we deal with weakly overlapping resonances. The results of the
fits are shown as solid lines in the lower two panels of
Fig.~\ref{fig3} and in the lower panel of Fig.~\ref{fig4}. For an
exponential decay in time, the curves in these panels should be
straight lines. This is clearly not the case. The solid lines in
the upper two panels of Fig.~\ref{fig3} and in the upper panel of
Fig.~\ref{fig4} are the Fourier transforms of the VWZ fits. In
Fig.~\ref{fig3} they agree well with the data points, 
up to small discrepancies which are attributed to finite--range--of--data errors.   
In the upper panel of Fig.~\ref{fig4} the discrepancy between fits and data
points is displayed more clearly than in the lower panel.

The quality of the agreement between data and fits in
Figs.~\ref{fig3} and~\ref{fig4} is assessed in terms of a highly
sensitive goodness--of--fit (GOF) test (the Fourier coefficients
scatter over more than five orders of magnitude!). The fit of
Eq.~(\ref{vwz_formel}) determines the expectation value $\overline
{x_t}$ of $x_t$ and, thus, $\eta_t = \ln \overline{x_t}$ in
Eq.~(\ref{vertlog}). If the distribution of the $y_t$ were
Gaussian, the GOF test would be defined in terms of $\sum_t (y_t -
\eta_t)^2$. The appropriate generalization for the distribution
$P(y_t)$ in Eq.~(\ref{vertlog}) is the expression $I \propto
\sum_t \left[ \exp{(y_t - \eta_t)}-(y_t - \eta_t) - 1 \right]$,
see Chaps.~14, 16 of Ref.~\cite{Har03}. This quantity is
non--negative and vanishes exactly if the data coincide with the
model for all $t$. For large $M$, $I$ is approximately
$\chi^2$--distributed with $M$ degrees of freedom. For each
frequency interval of length 1~GHz we have $M = 2400$, since each
of the three excitation functions $S_{11}(f)$, $S_{12}(f)$ and
$S_{22}(f)$ contributes 800 Fourier coefficients. We admit a 10~\%
probability for an erroneous decision. The fit using
Eq.~(\ref{vwz_formel}) is accepted in all intervals below
$f=10$~GHz and is rejected in all intervals but one above 10~GHz.
A similarly thorough and mathematically reliable test of the
theory of chaotic scattering has not been performed before, see
Refs.~\cite{vwz_exp,Sch03}. This fact motivated our work. We conclude
that Eq.~(\ref{vwz_formel}) is compatible with our data as long as the
resonator supports only two--dimensional modes and simulates a chaotic
billiard. We have numerically simulated the fluctuations above
10.1~GHz under the assumption that the two vertical modes do not
interact and the Hamiltonian matrix is block--diagonal, each block
taken from the GOE. In this way we reproduced qualitatively the
results of Fig.~\ref{fig4}. In the sense that the GOE describes full
chaos, a block--diagonal random matrix represents additional
symmetries. The disagreement between theory and experiment above
10.1 GHz shows that our test is sensitive to the existence of such
symmetries. We conclude that first, in the regime of overlapping resonances
our test is sensitive to symmetries in a Hamiltonian system and
second, that Eq.~(\ref{vwz_formel}) is compatible with the data as
long as the scattering system is fully chaotic.

{\it Summary}. We have investigated a chaotic microwave resonator in
the regime of weakly overlapping resonances $\Gamma \approx D$.  The
distributions of $S$--matrix elements are not Gaussian. In each of 13
frequency intervals we determined 2400 uncorrelated Fourier
coefficients of the elements of the scattering matrix.
Surprisingly, these have nearly Gaussian distributions. The data
were used to test the VWZ theory of chaotic scattering. The predicted
non--exponential decay in time of resonator modes and the frequency
dependence of the elastic enhancement factor are confirmed. Our
goodness--of--fit test is based on a large number of data points 
and constitutes the most sensitive test of the theory of quantum
chaotic scattering for weakly overlapping resonances performed so
far. We show that VWZ is compatible with the data as long as the
resonator simulates a fully chaotic quantum system. The theory
can, thus, be used with confidence to predict average cross sections
and $S$--matrix correlation functions. The agreement fails when a
second vertical mode appears. This suggests that our analysis may
serve as a tool to detect symmetries and/or regular motion within a
chaotic system in the regime of overlapping resonances.

\begin{acknowledgments}
We thank J.J.M. Verbaarschot for useful discussions, C. Lewenkopf
and A. M\"{u}ller for technical help, and M. Taheri Gelevarzi for
the construction of the microwave resonator. T.F. acknowledges a
fellowship from Studienstiftung des Deutschen Volkes and F.S. from
Deutsche Telekom Foundation. This work was supported by the DFG
within the SFB~634.
\end{acknowledgments}


\begin{thebibliography}{99}
\bibitem{footnote}The Gaussian distribution of the Fourier
coefficients does not contradict the non--Gaussian distribution of
the $S$--matrix elements: Moments of $\tilde{S}(t)$ are given by
integrals over correlation functions of $S(E)$ while moments of
$S(E)$ involve those same correlation functions at fixed values of
the arguments.
\end{thebibliography}
\end{document}